\documentclass[12pt]{article}
\usepackage{lineno}
\usepackage{epsfig}
\usepackage{amsmath}
\usepackage{hhline}
\usepackage{amssymb}
\usepackage{times}
\usepackage{cite}

\newif\ifpdf
\ifx\pdfoutput\undefined
    \pdffalse          
\else
    \pdfoutput=1       
    \pdftrue
\fi

\ifpdf
\usepackage{thumbpdf}
\usepackage[pdftex,
        colorlinks=true,
        urlcolor=rltblue,       
        filecolor=rltgreen,     
        linkcolor=rltred,       
        pdftitle={Search for Doubly-Charged Higgs Boson Production at HERA},
        pdfauthor={H1},
        pdfsubject={Template file},
        pdfkeywords={High-Energy Physics, Particle Physics},
        pdfpagemode=None,
        bookmarksopen=true]{hyperref}

\definecolor{rltred}{rgb}{0.75,0,0}
\definecolor{rltgreen}{rgb}{0,0.5,0}
\definecolor{rltblue}{rgb}{0,0,0.75}
 
\fi

\newlength{\dinwidth}
\newlength{\dinmargin}
\begin{document}  
\def\GeV{\hbox{$\;\hbox{\rm GeV}$}}

\newcommand{\pom}{{I\!\!P}}
\newcommand{\reg}{{I\!\!R}}
\newcommand{\slowpi}{\pi_{\mathit{slow}}}
\newcommand{\fiidiii}{F_2^{D(3)}}
\newcommand{\fiidiiiarg}{\fiidiii\,(\beta,\,Q^2,\,x)}
\newcommand{\n}{1.19\pm 0.06 (stat.) \pm0.07 (syst.)}
\newcommand{\nz}{1.30\pm 0.08 (stat.)^{+0.08}_{-0.14} (syst.)}
\newcommand{\fiidiiiful}{F_2^{D(4)}\,(\beta,\,Q^2,\,x,\,t)}
\newcommand{\fiipom}{\tilde F_2^D}
\newcommand{\ALPHA}{1.10\pm0.03 (stat.) \pm0.04 (syst.)}
\newcommand{\ALPHAZ}{1.15\pm0.04 (stat.)^{+0.04}_{-0.07} (syst.)}
\newcommand{\fiipomarg}{\fiipom\,(\beta,\,Q^2)}
\newcommand{\pomflux}{f_{\pom / p}}
\newcommand{\nxpom}{1.19\pm 0.06 (stat.) \pm0.07 (syst.)}
\newcommand {\gapprox}
   {\raisebox{-0.7ex}{$\stackrel {\textstyle>}{\sim}$}}
\newcommand {\lapprox}
   {\raisebox{-0.7ex}{$\stackrel {\textstyle<}{\sim}$}}
\def\gsim{\,\lower.25ex\hbox{$\scriptstyle\sim$}\kern-1.30ex%
\raise 0.55ex\hbox{$\scriptstyle >$}\,}
\def\lsim{\,\lower.25ex\hbox{$\scriptstyle\sim$}\kern-1.30ex%
\raise 0.55ex\hbox{$\scriptstyle <$}\,}
\newcommand{\pomfluxarg}{f_{\pom / p}\,(x_\pom)}
\newcommand{\dsf}{\mbox{$F_2^{D(3)}$}}
\newcommand{\dsfva}{\mbox{$F_2^{D(3)}(\beta,Q^2,x_{I\!\!P})$}}
\newcommand{\dsfvb}{\mbox{$F_2^{D(3)}(\beta,Q^2,x)$}}
\newcommand{\dsfpom}{$F_2^{I\!\!P}$}
\newcommand{\gap}{\stackrel{>}{\sim}}
\newcommand{\lap}{\stackrel{<}{\sim}}
\newcommand{\fem}{$F_2^{em}$}
\newcommand{\tsnmp}{$\tilde{\sigma}_{NC}(e^{\mp})$}
\newcommand{\tsnm}{$\tilde{\sigma}_{NC}(e^-)$}
\newcommand{\tsnp}{$\tilde{\sigma}_{NC}(e^+)$}
\newcommand{\st}{$\star$}
\newcommand{\sst}{$\star \star$}
\newcommand{\ssst}{$\star \star \star$}
\newcommand{\sssst}{$\star \star \star \star$}
\newcommand{\tw}{\theta_W}
\newcommand{\sw}{\sin{\theta_W}}
\newcommand{\cw}{\cos{\theta_W}}
\newcommand{\sww}{\sin^2{\theta_W}}
\newcommand{\cww}{\cos^2{\theta_W}}
\newcommand{\trm}{m_{\perp}}
\newcommand{\trp}{p_{\perp}}
\newcommand{\trmm}{m_{\perp}^2}
\newcommand{\trpp}{p_{\perp}^2}
\newcommand{\alp}{\alpha_s}

\newcommand{\alps}{\alpha_s}
\newcommand{\sqrts}{$\sqrt{s}$}
\newcommand{\LO}{$O(\alpha_s^0)$}
\newcommand{\Oa}{$O(\alpha_s)$}
\newcommand{\Oaa}{$O(\alpha_s^2)$}
\newcommand{\PT}{p_{\perp}}
\newcommand{\JPSI}{J/\psi}
\newcommand{\sh}{\hat{s}}
\newcommand{\uh}{\hat{u}}
\newcommand{\MP}{m_{J/\psi}}
\newcommand{\PO}{I\!\!P}
\newcommand{\xbj}{x}
\newcommand{\xpom}{x_{\PO}}
\newcommand{\ttbs}{\char'134}
\newcommand{\xpomlo}{3\times10^{-4}}  
\newcommand{\xpomup}{0.05}  
\newcommand{\dgr}{^\circ}
\newcommand{\pbarnt}{\,\mbox{{\rm pb$^{-1}$}}}
\newcommand{\gev}{\,\mbox{GeV}}
\newcommand{\WBoson}{\mbox{$W$}}
\newcommand{\fbarn}{\,\mbox{{\rm fb}}}
\newcommand{\fbarnt}{\,\mbox{{\rm fb$^{-1}$}}}
\newcommand{\pb}{\,\mbox{{\rm pb}}}
%
%
\newcommand{\qsq}{\ensuremath{Q^2} }
\newcommand{\gevsq}{\ensuremath{\mathrm{GeV}^2} }
\newcommand{\et}{\ensuremath{E_t^*} }
\newcommand{\rap}{\ensuremath{\eta^*} }
\newcommand{\gp}{\ensuremath{\gamma^*}p }
\newcommand{\dsiget}{\ensuremath{{\rm d}\sigma_{ep}/{\rm d}E_t^*} }
\newcommand{\dsigrap}{\ensuremath{{\rm d}\sigma_{ep}/{\rm d}\eta^*} }
\def\Journal#1#2#3#4{{#1} {\bf #2} (#3) #4}
\def\NCA{\em Nuovo Cimento}
\def\NIM{\em Nucl. Instrum. Methods}
\def\NIMA{{\em Nucl. Instrum. Methods} {\bf A}}
\def\NPB{{\em Nucl. Phys.}   {\bf B}}
\def\PLB{{\em Phys. Lett.}   {\bf B}}
\def\PRL{\em Phys. Rev. Lett.}
\def\PRD{{\em Phys. Rev.}    {\bf D}}
\def\ZPC{{\em Z. Phys.}      {\bf C}}
\def\EJC{{\em Eur. Phys. J.} {\bf C}}
\def\CPC{\em Comp. Phys. Commun.}

\newcommand{\mhpp}{\mbox{$M_H$}}
\newcommand{\hpp}{\mbox{$H^{\pm\pm}$}}

\begin{titlepage}

\noindent
\begin{flushleft}
DESY 06-038\hfill ISSN 0418-9833\\
April 2006
\end{flushleft}

\vspace{2cm}

\begin{center}
\begin{Large}

{\bf Search for Doubly-Charged Higgs Boson Production at HERA}

\vspace{2cm}

H1 Collaboration

\end{Large}
\end{center}

\vspace{2cm}

\begin{abstract}
\noindent
A search for the single production of doubly-charged Higgs bosons
$H^{\pm \pm}$ in $ep$ collisions is presented.
The signal is searched for via the Higgs decays 
into a high mass pair of same charge leptons, one of them being an electron.
The analysis uses up to $118$~pb$^{-1}$ of $ep$ data collected by the
H1 experiment at HERA.
No evidence for doubly-charged Higgs production is observed and
mass dependent upper limits are derived on the Yukawa couplings $h_{el}$ of the
Higgs boson to an electron-lepton pair.
Assuming that the doubly-charged Higgs only decays into
an electron and a muon via a coupling of electromagnetic
strength $h_{e \mu} = \sqrt{ 4 \pi \alpha_{em}} = 0.3$,
a lower limit of $141$~GeV on the $H^{\pm\pm}$ mass is obtained
at the $95\%$ confidence level.
For a doubly-charged Higgs decaying only into an electron and a tau 
and a coupling $h_{e\tau} = 0.3$, masses below $112$~GeV are ruled out.

\end{abstract}

\vspace{1.5cm}

\begin{center}
Submitted to {\it Phys. Lett. B}
\end{center}

\end{titlepage}

\newpage


\noindent
A.~Aktas$^{9}$,                
V.~Andreev$^{25}$,             
T.~Anthonis$^{3}$,             
B.~Antunovic$^{26}$,           
S.~Aplin$^{9}$,                
A.~Asmone$^{33}$,              
A.~Astvatsatourov$^{3}$,       
A.~Babaev$^{24, \dagger}$,     
S.~Backovic$^{30}$,            
A.~Baghdasaryan$^{37}$
% ...

\newpage
\bigskip{\noindent \it
 $ ^{1}$ I. Physikalisches Institut der RWTH, Aachen, Germany$^{ a}$ \\
% ...
$ ^{3}$ Inter-University Institute for High Energies ULB-VUB, Brussels;
          Universiteit Antwerpen, Antwerpen; Belgium$^{ c}$ \\
% ...
$ ^{9}$ DESY, Hamburg, Germany \\
% ...
$ ^{24}$ Institute for Theoretical and Experimental Physics,
          Moscow, Russia$^{ k}$ \\
 $ ^{25}$ Lebedev Physical Institute, Moscow, Russia$^{ e}$ \\
 $ ^{26}$ Max-Planck-Institut f\"ur Physik, M\"unchen, Germany \\
% ...
$ ^{30}$ Faculty of Science, University of Montenegro,
          Podgorica, Serbia and Montenegro$^{ e}$ \\
% ...
$ ^{33}$ Dipartimento di Fisica Universit\`a di Roma Tre
          and INFN Roma~3, Roma, Italy \\
% ...
$ ^{37}$ Yerevan Physics Institute, Yerevan, Armenia \\
% ...
$ ^{\dagger}$ Deceased \\

\bigskip \noindent
 $ ^a$ Supported by the Bundesministerium f\"ur Bildung und Forschung, FRG,
      under contract numbers 05 H1 1GUA /1, 05 H1 1PAA /1, 05 H1 1PAB /9,
      05 H1 1PEA /6, 05 H1 1VHA /7 and 05 H1 1VHB /5 \\
% ...
$ ^c$ Supported by FNRS-FWO-Vlaanderen, IISN-IIKW and IWT
      and  by Interuniversity
Attraction Poles Programme,
      Belgian Science Policy \\
% ...
$ ^e$ Supported by the Deutsche Forschungsgemeinschaft \\
% ...
$ ^k$ Partially Supported by Russian Foundation
      for Basic Research,  grants  03-02-17291
      and  04-02-16445 \\
}

\section{Introduction}

\noindent
Doubly-charged Higgs bosons ($H^{\pm \pm}$) appear when
the Higgs sector of the Standard Model (SM) is extended by one or more
triplet(s) with non-zero hypercharge~\cite{HTM,pati,moha1}.
Examples are provided by some Left-Right Symmetric models~\cite{moha3},
or their supersymmetric extensions, which are of particular
interest since they provide a mechanism to generate small non-zero neutrino masses.
Such models can lead to a doubly-charged Higgs boson light enough~\cite{LIGHT} to be
produced at the existing colliders.
The Higgs triplet(s) may be coupled to matter fields
via Yukawa couplings which are generally not related to the
fermion masses.
A non-vanishing coupling of a doubly-charged Higgs to an electron would
allow its single production in $ep$ collisions at HERA. This possibility is
investigated in this paper with a search for doubly-charged Higgs bosons decaying
into a high mass pair of same charge leptons, one of them being an electron.

\noindent
An analysis of multi-electron events was already
presented by the H1 collaboration~\cite{me}. 
Six events were observed with a di-electron mass
above $100$~GeV, a domain in which the Standard Model prediction
is low. 
In the present paper the compatibility of these events with the hypothesis of a doubly-charged Higgs
coupling to $ee$ is addressed and a further search for a \hpp\ boson coupling to $e\mu$ and
$e\tau$ is performed.
The analysis is based on $ep$ data collected by the H1 experiment between
$1994$ and $2000$, which amount to a luminosity of up to $118$~pb$^{-1}$.

\section{Phenomenology} 

At tree level, doubly-charged Higgs bosons couple
only to charged leptons and to other Higgs and gauge bosons.
Couplings to quark pairs are forbidden by charge conservation.
The couplings of a doubly-charged Higgs to charged leptons can
be generically described by the Lagrangian:
\begin{equation}
 {\cal{L}} =  \sum_{i,j} h^{L,R}_{l_{i} \, l_{j}} \, H_{L,R}^{++} \, \bar{\l_i}^c \, P_{L,R} \, \l_j \qquad {\mbox{  + \ h.c.}} \, \, \, ,
 \label{eq:lag}
\end{equation}
where $l$ are the charged lepton fields, 
$l^c$ denote the charge conjugate fields,
$i,j$ are generation indices, and
$P_{L,R} = (1 \mp \gamma_5)/2$. 
The Higgs fields $H_{L,R}^{++}$ coupling to left-handed or right-handed leptons correspond to
different particles and not all models predict their simultaneous existence.
The Yukawa couplings $h^{L,R}_{l_{i}l_{j}} = h^{L,R}_{l_{j}l_{i}}$ are free parameters of the model.

%
%
\begin{figure}[htb]
  \begin{center}
 \begin{tabular}{ccc}
  \epsfig{file=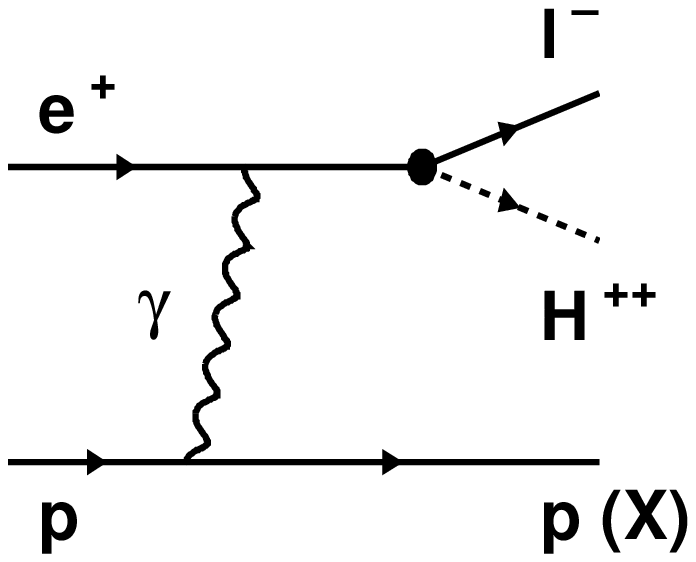,width=5.cm} &
  \epsfig{file=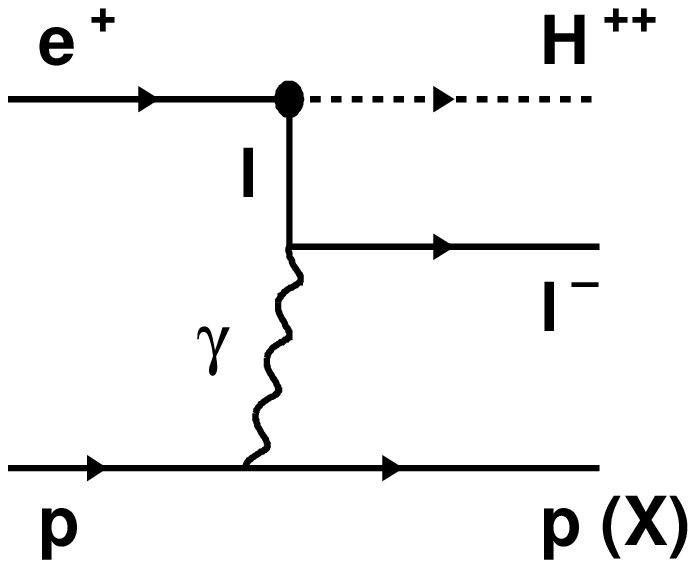,width=5.cm} &
  \epsfig{file=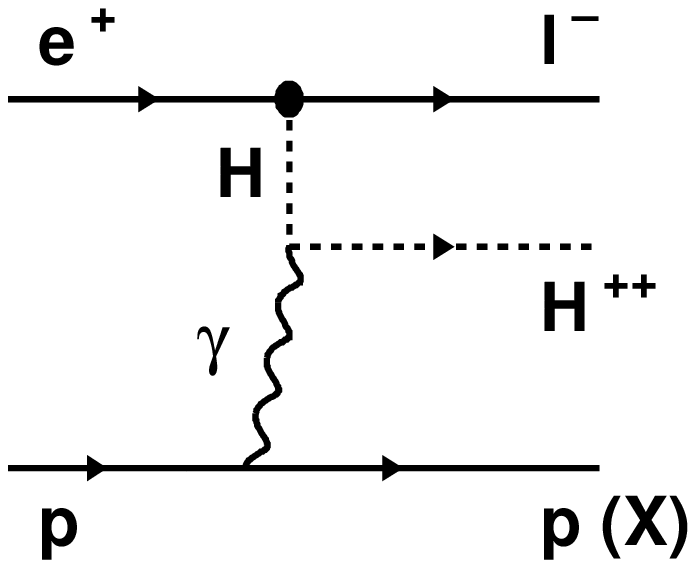,width=5.cm}
 \end{tabular}
  \end{center}
  \caption{
Diagrams for the single production of a doubly-charged Higgs boson
in $e^{+}p$ collisions at HERA via the $h_{el}$ coupling.
The hadronic final state is denoted by $p$ ($X$)
in the elastic (inelastic) case, where the initial
proton remains intact (dissociates).
The contribution of $Z$ exchange can be safely neglected.
}
\label{fig:feyn}
\end{figure}

\noindent
The phenomenology of doubly-charged Higgs production
at HERA was first discussed
in~\cite{zeus}.
For a non-vanishing coupling $h_{e\l}^{L,R}$ the single production
of a doubly-charged Higgs boson is possible at HERA
in $e\gamma^*$ interactions
via the diagrams shown in Fig.~\ref{fig:feyn}, 
where a photon is radiated off the proton  or one of its
constituent quarks.
The proton may remain intact or be broken during this interaction,
leading to an elastic or inelastic reaction, respectively.
With longitudinally unpolarised lepton beams, as were delivered
by HERA until $2000$, 
the \hpp\ production cross section does not depend 
on whether the Higgs couples to left-handed
or right-handed leptons.
Hence a generic case is considered here of a
doubly-charged Higgs boson
which couples to either left-handed or right-handed leptons and the $L,R$ indices
are dropped in the following. 

\noindent
Within the mass range considered in this analysis, it is assumed that 
decays of the $H^{\pm \pm}$ into gauge bosons and other Higgs particles
are not allowed kinematically such that the doubly-charged Higgs only decays
via its Yukawa couplings into a lepton pair. 
 
\noindent
Indirect upper bounds on the Yukawa couplings of a doubly-charged
Higgs to leptons are reviewed in~\cite{indlim1}.
The coupling $h_{ee}$ of a doubly-charged Higgs to an electron pair is 
constrained by the contribution of virtual $H^{\pm\pm}$ exchange
to Bhabha scattering in $e^+e^-$ collisions. A recent OPAL 
analysis~\cite{opal-singleprod} sets the constraint
$h_{ee} < 0.14$ for a doubly-charged Higgs mass  $M_H = 100 \GeV$.
From low energy $e^+ e^-$ data, coupling values of ${\cal{O}} (0.1)$ 
are allowed for $h_{e \mu}$ and $h_{e \tau}$
for a Higgs mass of $100 \GeV$~\cite{DAVIDSON}.
Taking these indirect constraints into account, the production of a  
doubly-charged Higgs mediated by $h_{ee}$, $h_{e \mu}$ or $h_{e \tau}$ might
be observable at HERA.
The Higgs signal would manifest itself as a peak in the invariant mass distribution 
of same charge $ee$, $e \mu$ or $e \tau$ leptons, respectively.
For the range of masses and couplings
probed in this analysis, the Higgs decay length is vanishingly small
but its width remains negligible compared
to the experimental resolution on the mass of the lepton pair.

\section{Simulation of the Signal and Standard Model Backgrounds}

The calculation of the cross section for doubly-charged Higgs production, as well
as the simulation of signal events, relies on a dedicated Monte Carlo
program developed for this analysis. 
The differential cross sections are integrated using
the VEGAS package~\cite{VEGAS}.
Different approaches are followed
depending on the photon virtuality $Q^2$ and on the mass $W$ of the hadronic
final state:
\begin{itemize}
\item in the inelastic region ($W > m_p + m_{\pi}$, with the
      proton mass $m_p$ and the pion mass $m_{\pi}$) and when the photon virtuality
      is large ($Q^2 > 4 \GeV^2$), the interaction involves a quark inside
      the proton.
      The squared amplitude of the process $e^{\pm} q \rightarrow e^{\mp} H^{\pm \pm} q$
      is evaluated using the CompHEP package~\cite{comphep,ROMANENKO}.
      The parton densities in the proton are taken
      from the CTEQ4L~\cite{cteq} parameterisation and are evaluated at
      the scale $\sqrt{Q^2}$.
      The parton shower approach~\cite{pythia} based
      on the DGLAP~\cite{DGLAP} evolution equations is applied
      to simulate QCD corrections in the initial and final states, and the hadronisation
      is performed using PYTHIA~6.1~\cite{pythia}.
\item for the elastic region ($W$ = $m_p$) and the inelastic region at low $Q^2$
      ($W > m_p + m_{\pi}$, $Q^2 < 4 \GeV^2$),
      the squared amplitude 
      is calculated using the FORM program~\cite{FORM}.
      The hadronic tensor is parameterised in terms of the usual
      electromagnetic structure functions $F_1 (x, Q^2)$ and $F_2(x, Q^2)$ of the proton,
      where $x = Q^2 / (W^2 + Q^2 - m^2_p)$.
      For the elastic process these structure functions 
      are expressed in terms
      of the electric and magnetic form factors 
      of the proton.
      For the low $Q^2$ inelastic region 
      they are taken from analytical parameterisations~\cite{BRASSE}.
      The simulation of the hadronic final state for low $Q^2$ inelastic
      events is performed 
      via an interface to the SOPHIA program~\cite{SOPHIA}.
\end{itemize}

\noindent
For a Yukawa coupling $h_{ee}$ or $h_{e \mu}$ of electromagnetic strength 
($h = \sqrt{4 \pi \alpha_{em}} = 0.3$)
the total cross section amounts to $0.39$~pb ($0.04$~pb) for a Higgs mass of
$100$~GeV ($150$~GeV).
The low $Q^2$ (high $Q^2$) inelastic contribution is found
to be $\sim 30 \%$ ($\sim 20 \%$)
of the total cross section in the mass range $80-150$~GeV.
The cross section for producing a doubly-charged Higgs via a
coupling $h_{e \tau}$ is lower by about $40\%$ due to the 
non-negligible mass of the $\tau$ lepton produced in association with
the Higgs.

\noindent
The theoretical uncertainty on the cross sections obtained is
taken to be $4\%$ in the mass range considered.
This is derived from an assessed
uncertainty of $2\%$ on the proton form factors~\cite{SLAC} and
from the uncertainty on the scale at which the parton densities for the inelastic
contribution are evaluated. 
The latter uncertainty is estimated from the variation of the computed cross section
as this scale is changed from ${\sqrt {Q^2}} /2$ to $2 {\sqrt {Q^2}}$.

\noindent
Separate signal event samples corresponding to the production and decay of a doubly-charged
Higgs via a coupling $h_{ee}$, $h_{e \mu}$ and $h_{e \tau}$ have been
produced for Higgs masses ranging between $80$ and $150 \GeV$, in steps of $10 \GeV$.

\noindent
Di-electron production, which proceeds mainly via two-photon interactions,
constitutes an irreducible background for $ee$ final
states. The production of muon or tau pairs constitutes a background
for the $e \mu$ and $e \tau$ analyses when the scattered electron is
detected.
The Cabibbo-Parisi process $ee \rightarrow \gamma, Z \rightarrow ll$, in which the 
incoming electron interacts with an
electron emitted from a photon radiated from the proton,
contributes at high transverse momentum only.
The Drell-Yan process was calculated in~\cite{dy} and 
found to be negligible.
All these processes are simulated using the
GRAPE Monte Carlo generator~\cite{grape}, which also takes into
account contributions from Bremsstrahlung with subsequent
photon conversion into a lepton pair
and electroweak contributions.

\noindent
Experimental backgrounds come dominantly from Neutral Current Deep Inelastic
Scattering (NC DIS) where a jet is misidentified as an electron, a muon
or a tau. Compton scattering is also a source of background for $ee$ final
states when the photon is misidentified as an electron.
These processes are simulated with the DJANGO~\cite{django}
and WABGEN~\cite{wabgen} generators.

\noindent
All generated events are passed through the full simulation of the H1 apparatus
and are reconstructed using the same program chain as for the data.

\section{The H1 Detector}

A detailed description of the H1 experiment can be found in \cite{Abt:1996xv}.
Only the H1 detector components relevant to the
present analysis are briefly described here.
Jets and electrons are measured with the Liquid
Argon (LAr) calorimeter~\cite{Andrieu:1993kh}, which covers the polar angle\footnote{ 
  The origin of the H1 
  coordinate system is the nominal $ep$ interaction point, with 
  the direction of the proton beam defining the positive 
  $z$-axis (forward region). The transverse momenta are measured 
  in the $xy$ plane. 
  The 
  pseudorapidity $\eta$ is related to the polar 
  angle $\theta$ by $\eta = -\ln \, \tan (\theta/2)$.} range
$4^\circ < \theta < 154^\circ$.
Electromagnetic shower energies are measured with a precision of
$\sigma (E)/E = 12\%/ \sqrt{E/\mbox{GeV}} \oplus 1\%$ and hadronic energies
with $\sigma (E)/E = 50\%/\sqrt{E/\mbox{GeV}} \oplus 2\%$, as determined
in test beams~\cite{h1calotestbeams}.
In the backward region a 
lead/scintillating-fibre\footnote{
Before $1995$ a lead-scintillator calorimeter was used.
} (SpaCal) calorimeter~\cite{spacal} covers the range $155^\circ < \theta < 178^\circ$.
The central ($20^\circ < \theta < 160^\circ$) and forward ($7^\circ < \theta < 25^\circ$)
tracking detectors are used to
measure charged particle trajectories, to reconstruct the interaction
vertex and to supplement the measurement of the hadronic energy.
The LAr and inner tracking detectors are enclosed in a super-conducting magnetic
coil with a strength of $1.15$~T.
The return yoke of the coil is the outermost part of the detector and is
equipped with streamer tubes forming the central muon detector
($4^\circ < \theta < 171^\circ$).
In the forward region of the detector ($3^\circ < \theta < 17^\circ$) a set of
drift chamber layers (the forward muon system) detects muons and, together with an
iron toroidal magnet, allows a momentum measurement.
The luminosity measurement is based on the Bethe-Heitler process  $ep \rightarrow ep \gamma$,
where the photon is detected in a calorimeter located
downstream of the interaction point.


\section{Data Analysis}

The analyses of $ee$ and $e \mu$ final states use the full $e^{\pm}p$ data set recorded
in the period $1994$-$2000$, which corresponds to an integrated luminosity
of $118 \pb^{-1}$.
The analysis of $e \tau$ final states makes use of the
$e^+ p$ data collected in the years $1996$-$1997$ and $1999$-$2000$, which amount
to a luminosity of $88 \pb^{-1}$.
The HERA collider was operated at a centre-of-mass energy $\sqrt{s}$ of $300$~GeV
in $1994$-$1997$ and of $318$~GeV in $1998$-$2000$.
 
\noindent
Events are first selected by requiring that the longitudinal position of
the vertex be within $35$~cm around the nominal interaction point.
In addition topological filters and timing vetoes are applied
to remove background events induced by
cosmic showers and other non-$ep$ sources.
The main triggers for the events are provided by the LAr calorimeter
and the muon system.

\subsection{Lepton Identification}

\label{section:id}

An electron\footnote{Unless otherwise stated,
the term ``electron'' is used in this paper
to generically describe electrons or positrons.}
candidate is identified by the presence of a compact and isolated
electromagnetic energy deposit above $5 \GeV$ in the LAr or SpaCal calorimeter.
The energy of the electron candidate is measured from the calorimetric information.
In the angular range $20^\circ < \theta < 150^\circ$ the electron identification
is complemented by tracking conditions, in which case the direction of the
electron candidate is given by that of the associated track. 
Electron candidates in the forward region, 
$5^\circ < \theta < 20^\circ$, are required to have an energy above $10 \GeV$.

\noindent
A muon candidate is identified by associating an isolated track in the
forward muon system or in the inner tracking system 
with a track segment or an energy deposit in the instrumented iron.
The muon momentum is measured from the track curvature in the toroidal
or solenoidal magnetic field, respectively.

\noindent
Tau leptons are preselected as described in~\cite{ref:simon} by requiring a track with transverse
momentum above $5 \GeV$ measured in the inner tracking detector.
The leptonic tau decays $\tau \rightarrow e \nu \nu$ and
$\tau \rightarrow \mu \nu \nu$ are reconstructed by matching the
selected track to an identified electron or muon.
Tracks that are not identified as electrons or muons 
are attributed to hadronic tau 
decays if at least $40\%$ of the track momentum is reconstructed in
the LAr calorimeter as matched clustered energy. 
In that case it is moreover required that the track belong to a narrow
jet: no other track should be reconstructed within $0.15<R<1.5$ around the
track direction, where $R=\sqrt{\Delta \eta^2 + \Delta \varphi^2}$
with $\Delta \eta$ and $\Delta \varphi$ being the distances in pseudorapidity
and azimuthal angle, respectively. 
The transverse momentum and the direction of the $\tau$ candidate 
are approximated
by those of the associated track.

\subsection{Analysis of the \boldmath{$H \rightarrow ee$} Decay}

\label{section:eemumu}

\begin{figure}[tb]
  \begin{center}
\begin{picture}(10,100)(0,40)
%
\put(-78.,76.){\epsfig{file=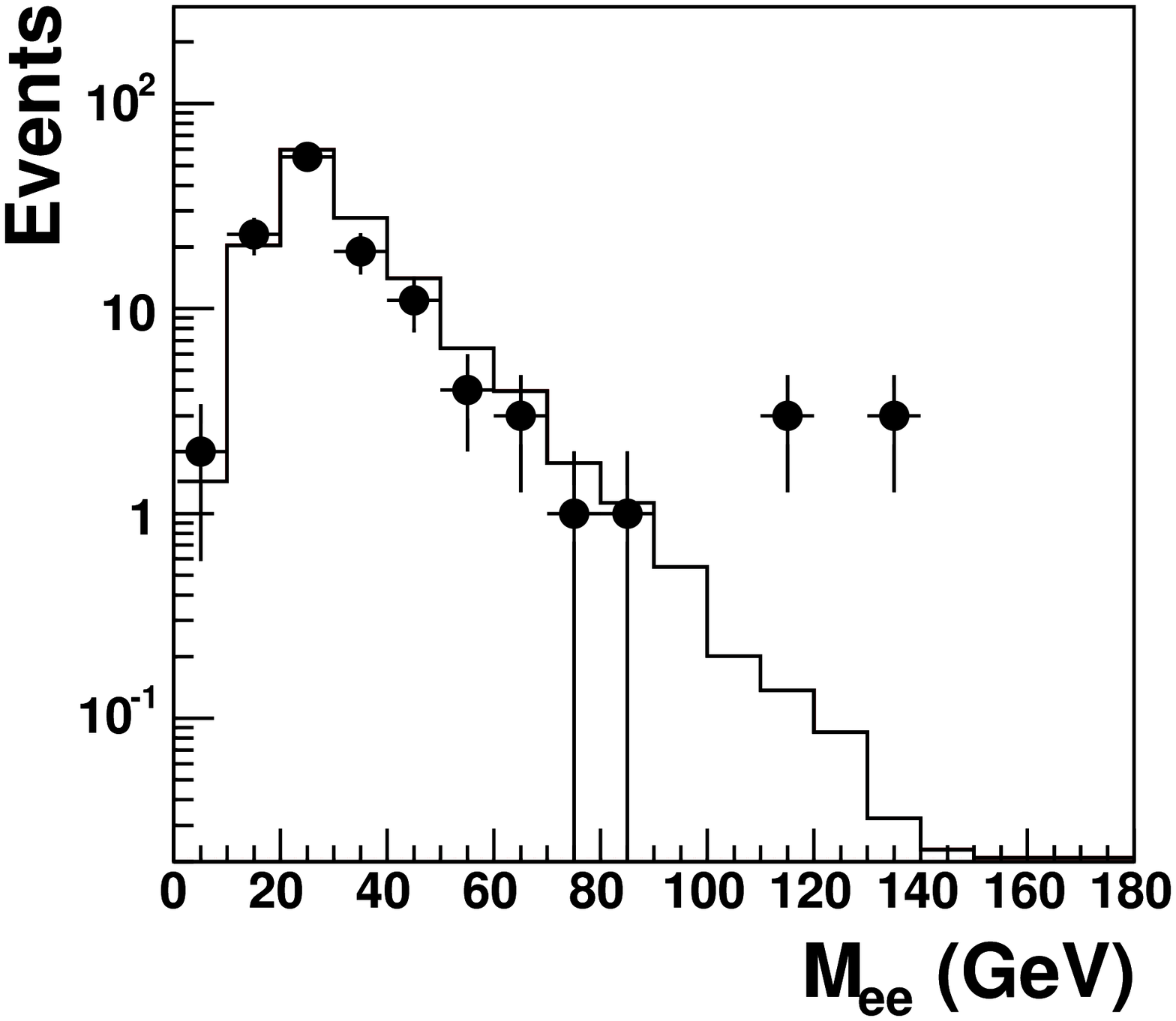,width=7.6cm,clip=}}
\put(-15.,133.){{\large \bf (a)}}
\put(5.,66.){\epsfig{file=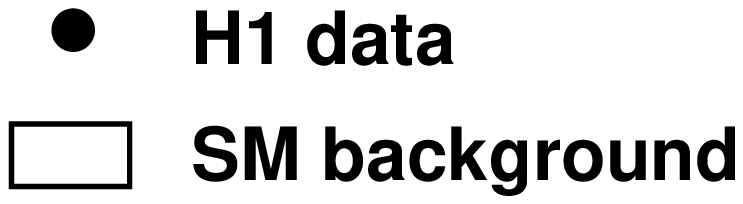,width=7.6cm,clip=}}
\put(-78.,5.){\epsfig{file=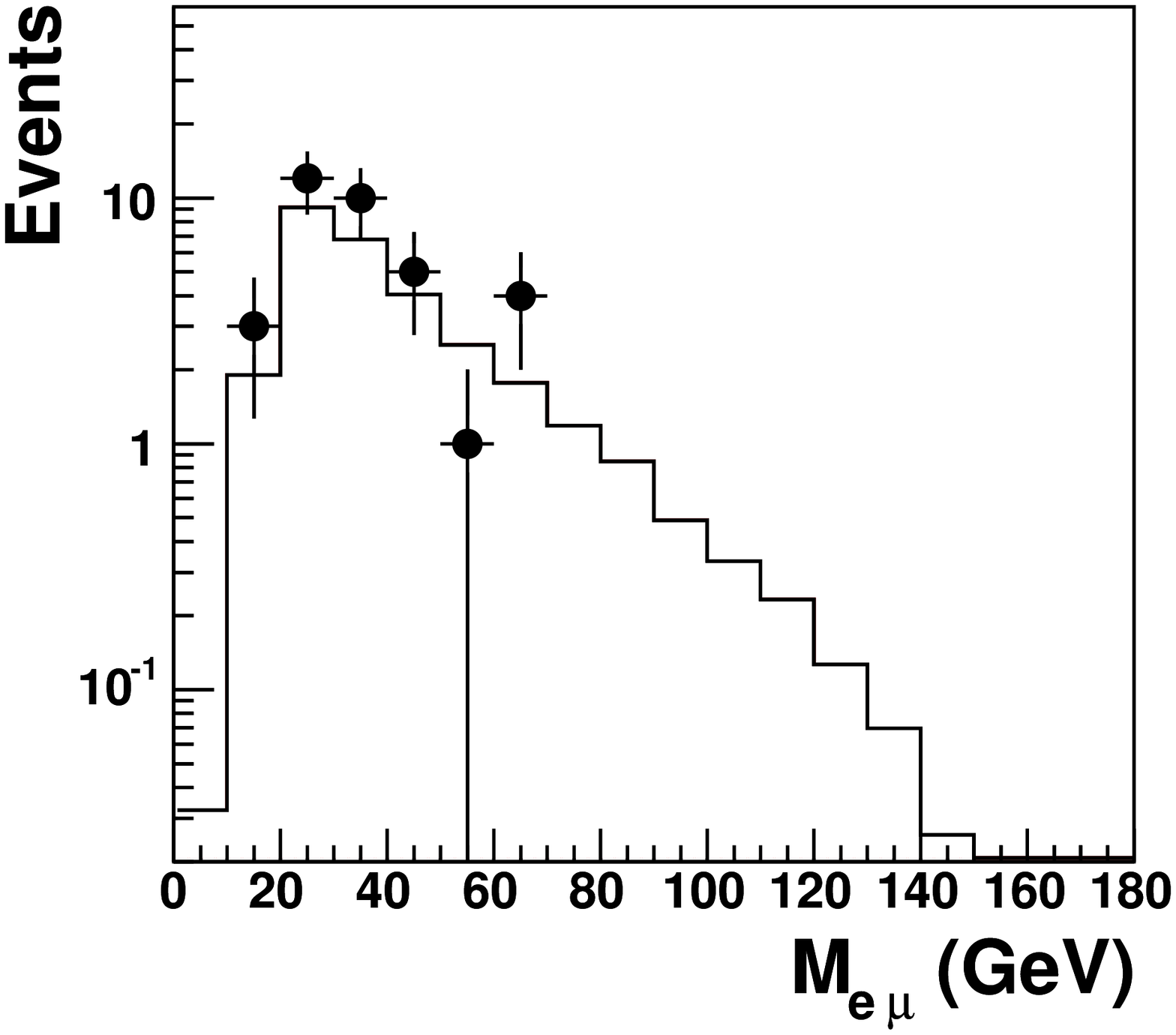,width=7.6cm,clip=}}
\put(-15.,62.){{\large \bf (b)}}
\put(5.,5.){\epsfig{file=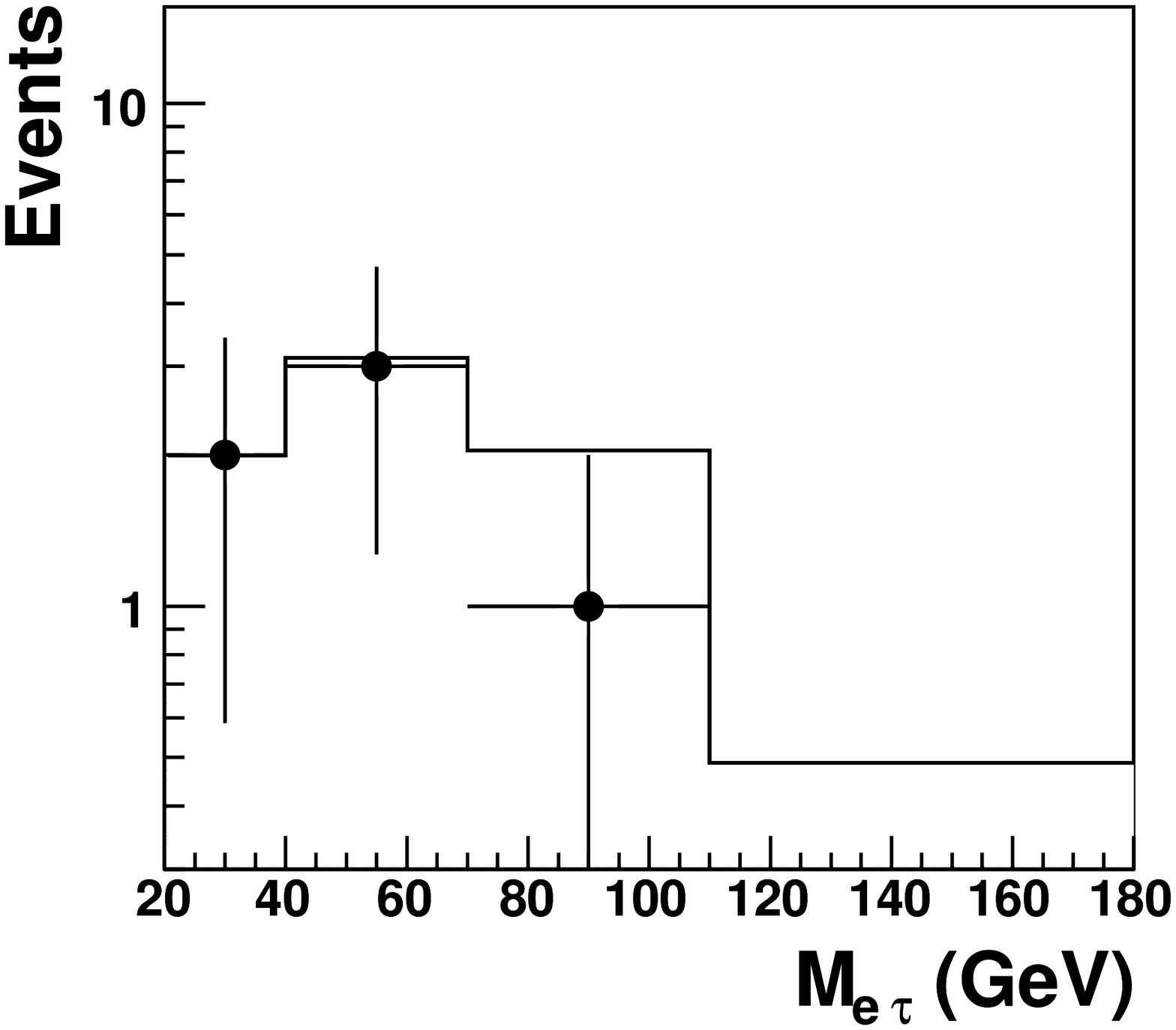,width=7.6cm,clip=}}
\put(68.,62.){{\large \bf (c)}}

\end{picture}
  \end{center}
\vspace*{3cm}
  \caption{
Distribution of (a) the invariant mass $M_{ee}$ of the two highest $P_T$ electrons
for multi-electron events,
(b) the electron-muon invariant mass $M_{e\mu}$, and (c) the electron-tau
candidate invariant mass $M_{e\tau}$.
The data (symbols) are compared with the Standard Model expectation (histogram).
The distributions are shown at the preselection level (see text).
}
\label{invmass}
\end{figure}

This analysis is based on the published H1 measurement of multi-electron
production~\cite{me}.
The event selection requires at least
two central ($20^{\circ} < \theta^{e} < 150^{\circ}$)
electron
candidates, one of them with a transverse momentum 
$P_T^{e1}  > 10$~GeV 
(ensuring a trigger efficiency close to $100\%$~\cite{Adloff:2003uh})
and the other one with $P_T^{e2} > 5$~GeV.
After this preselection, $125$ events are observed, in good
agreement with the SM expectation of $137.4 \pm 10.7$.
In each event, the two highest $P_T$ electrons, one of those being possibly 
outside the central region, are assigned to the Higgs candidate.
The distribution of their invariant mass $M_{ee}$
is shown in Fig.~\ref{invmass}a.
At low mass a good agreement is observed between data and the SM expectation
which is largely dominated by $\gamma \gamma$ contributions.
Six events are observed at $M_{ee} > 100$~GeV,
compared to the SM expectation of $0.53 \pm 0.08$.

\noindent
Further selection criteria are then applied, which are designed
to maximise the sensitivity of the analysis
to a possible \hpp\ signal.   
The charge measurement of the two leptons
assigned to the Higgs candidate is exploited.
In $e^+ p$ ($e^- p$) collisions, where $H^{++}$ ($H^{--}$) bosons 
could be produced,
events in which one of the two leptons is reliably assigned
a negative (positive) charge are rejected.
The charge assignment requires that the curvature
$\kappa$ of the track associated with the lepton be measured
with an error $\delta \kappa$ satisfying      
$\mid \kappa / \delta \kappa \mid \, > 2$.
The precise calorimetric measurement of the electron transverse momenta 
is further exploited by applying
an additional $M_{ee}$ dependent cut on the sum of the
transverse momenta of the two electrons assigned to the Higgs candidate.
The lower bound is optimised to  keep $95\%$ of the signal and
varies between $45$~GeV and $120$~GeV.
This cut suppresses events coming from $\gamma \gamma$ processes.
The efficiency for selecting signal events
varies from $50\%$ for a \hpp\ mass of $80$~GeV to
$35\%$ for a \hpp\ mass of $150$~GeV.
In this mass range the resolution on $M_{ee}$ varies between $2.5$~GeV and $5$~GeV.

\noindent
After these requirements, $3$ events are observed at $M_{ee} > 65 \GeV$,
in agreement with the SM expectation of
$2.45 \pm 0.11$ events. Amongst the six events\footnote{Out of these,
three do not fullfill the $M_{ee}$ dependent $P_T$ cut,
and two do not satisfy the charge requirement.} at $M_{ee} > 100 \GeV$,
only one satisfies the final selection criteria.

%
%

\subsection{Analysis of the \boldmath{$H \rightarrow e \mu$} Decay}

Events having one electron and one muon with minimal transverse momenta of
$P_T^{e} > 10$~GeV and $P_T^{\mu} > 5$~GeV are selected.
The polar angle of electron candidates
is restricted to $20^\circ < \theta^e <140^\circ$ to reduce the large background
arising from NC DIS events.
The $\theta$ range for muon candidates extends towards low angles,
$10^\circ < \theta^{\mu} < 140^\circ$, which
increases the efficiency for high \hpp\ masses.
The minimum transverse momentum required for electron candidates ensures a trigger
efficiency close to $100\%$ for these events.
After this preselection, $35$ data events are observed compared to a SM
expectation
of $29.6$ $\pm$ $3.4$. The distribution of the invariant mass of the
electron and
the muon $M_{e\mu}$ is shown in Fig.~\ref{invmass}b.
A good agreement is observed between the data and the SM expectation,
which is dominated by $\gamma \gamma$ contributions.

\noindent
For the final selection of $H \rightarrow e \mu$ candidates the charge
of the $e$ and $\mu$ is exploited using the same criteria as
used in section~\ref{section:eemumu}.
The efficiency for selecting signal events varies from $55\%$ to $40\%$
for a \hpp\ mass between $80$~GeV and $150$~GeV.
The resolution on $M_{e\mu}$ varies between $3$~GeV and $8$~GeV.
For $M_{e \mu} > 65 \GeV$ one event is observed while
$4.17 \pm 0.44$ events are expected from the SM.

\subsection{Analysis of the \boldmath{$H \rightarrow e \tau$} Decay}

The search for a $H^{++}$ boson decaying into $e\tau$ is performed in
three final states, depending on whether the $\tau$ decays into an
electron, a muon or hadronically ($h$). 
Details of this analysis can be found in~\cite{ref:simon}.
Events are selected which contain either two electrons ($ee$), or an 
electron and a muon ($e \mu$),
or an electron and a hadronic $\tau$ candidate ($eh$) as defined in section~\ref{section:id}.
The two leptons, or the electron and the hadronic-$\tau$ candidate, should have a transverse
momentum above $5 \GeV$, be in the angular range
$20^{\circ} < \theta < 140^{\circ}$, and be
separated from each other by $R > 2.5$ in pseudorapidity-azimuth.
One of them must have a transverse
momentum above $10 \GeV$, which ensures a trigger efficiency above
$95 \%$ in all three classes.  For events in the $e \mu$ class
the polar angle of the electron candidate is required to be below $120^{\circ}$.

\noindent
A significant amount of missing transverse and longitudinal momentum
is expected due to the neutrinos produced in the $\tau$ decays.
Events in the $ee$ class are required to have a missing transverse
momentum $P_T^{miss} > 8 \GeV$.
For the $eh$ class, which suffers from a large NC DIS background,
it is required that $P_T^{miss}>11 \GeV$, that the energy deposited
in the SpaCal calorimeter be below $5 \GeV$, and that
the variable $\sum_i E^i - P_z^i$,
where the sum runs over all visible particles, be smaller
than $49 \GeV$. For fully contained events $\sum_i E^i - P_z^i$ is expected
to peak at twice the lepton beam energy $E_0 = 27.5 \GeV$, i.e. $55$~GeV, while signal
events are concentrated at lower values due to the non observed neutrinos.
In total $6$ events are preselected, in agreement with the SM prediction of $7.8 \pm 1.5$.

\noindent
In each class, the $e \tau$ invariant mass $M_{e \tau}$ is reconstructed
by imposing longitudinal momentum and energy conservation, and by minimising 
the total momentum imbalance in the transverse plane. 
Tau leptons are assumed to decay with a vanishing
opening angle.
This method yields a resolution of about $4 \GeV$ on the mass $M_{e \tau}$.
Figure~\ref{invmass}c shows the $e \tau$ invariant mass distribution of the
selected events together with the SM expectation.

\noindent
For the final selection, events are rejected if the track associated with one of the
Higgs decay product candidates is
reliably assigned a negative charge, opposite to that of the incoming lepton beam.
The signal efficiencies depend only weakly on \mhpp.
The fractions of simulated $H \rightarrow e \tau$ events
which are reconstructed in the various classes are given in
table~\ref{tab:etau_cutflow}, for an example mass of $\mhpp=100$~GeV.
The total efficiency on the signal amounts to about $25 \%$.

\noindent
The final event yields
are also shown in table~\ref{tab:etau_cutflow}.
Only one event (in the $eh$ class) satisfies the final criteria, while
$2.1 \pm 0.5$ events are expected.

\begin{table}[htb]
 \begin{center}
\begin{tabular}{|c||c|c|c|}
   \hline
 Event           
               & \multicolumn{3}{|c|}{ $H^{++} \rightarrow e^+  \tau^+ $ final selection}
                \\ \cline{2-4}
 class & $N_{obs}$ & $N_{bckg}$ & {Signal fraction} \\ \hline \hline
   $e \mu$            & 0 & 0.27$\pm$ 0.02 & 6 \% \\ \hline
   $e h$         & 1 & 1.66 $\pm$ 0.48   & 12 \%  \\ \hline
   $ee$               & 0 & 0.14$\pm$ 0.04 &  7 \% \\ \hline \hline
   total                      & 1 & 2.07$\pm$ 0.54     & 25 \% \\ \hline
\end{tabular} 
 \caption[]
          {\label{tab:etau_cutflow}
            Number of observed ($N_{obs}$) and expected ($N_{bckg}$)
            events in each event class which satisfy all criteria to
            select $H^{++} \rightarrow e^+ \tau^+$ candidates
            with a mass $M_{e \tau}>65$~GeV.
            The last column shows the fractions of the
            $H \rightarrow e \tau$ Monte Carlo events which are
            reconstructed in the various classes, for a
            mass of $100$~GeV.
          }
 \end{center}
\end{table}

\subsection{Systematic Uncertainties}

The systematic uncertainties attributed to the Monte Carlo predictions
for the $ee$ analysis are detailed in~\cite{me}. 
The dominant systematic uncertainty is due
to the electron-track association efficiency, which is $90\%$ on average with an
uncertainty increasing
with decreasing polar angle from $3\%$ to $15\%$.
Systematic errors due to the uncertainty on the
electromagnetic energy scale (known at the
$0.7\%$ to $3\%$ level in the central and forward regions of the LAr calorimeter, respectively)
and on the trigger efficiency ($3\%$) are also
taken into account.

\noindent
For the $e \mu$ analysis, the dominant additional systematic
uncertainty is due to the muon identification efficiency known within 
$6 \%$~\cite{mm}. The uncertainty due to the reconstruction efficiency of
the central tracking detector for central muons contributes an additional
$3 \%$. 
The muon momentum scale is known within $5\%$,
and the trigger efficiency for $e \mu$ final
states is known within $3\%$.

\noindent
The same systematic uncertainties affect the SM expectations 
in the $ee$ and $e \mu$ classes of the $e \tau$ analysis. 
The uncertainty of the hadronic energy scale in the
LAr calorimeter ($4 \%$) constitutes another source of uncertainty
due to the cuts applied on the $P_T^{miss}$ and $\sum_i E -P_z$
variables.
For the $eh$ event class the dominant uncertainties on the SM
expectation, coming mainly from NC DIS processes, are due to the
uncertainty of $3\%$ of the track efficiency, to that of the
hadronic energy scale, and to that of the hadronisation model.

\noindent
The luminosity measurement leads to a normalisation
uncertainty of $1.5\%$.

\noindent
For both the expected signal and the predicted background, the
systematic uncertainties resulting from the sources listed above are added in quadrature.

\section{Interpretation}

%
%
\begin{figure}[htb]
  \begin{center}
  \begin{tabular}{cc}
%
 \epsfig{file=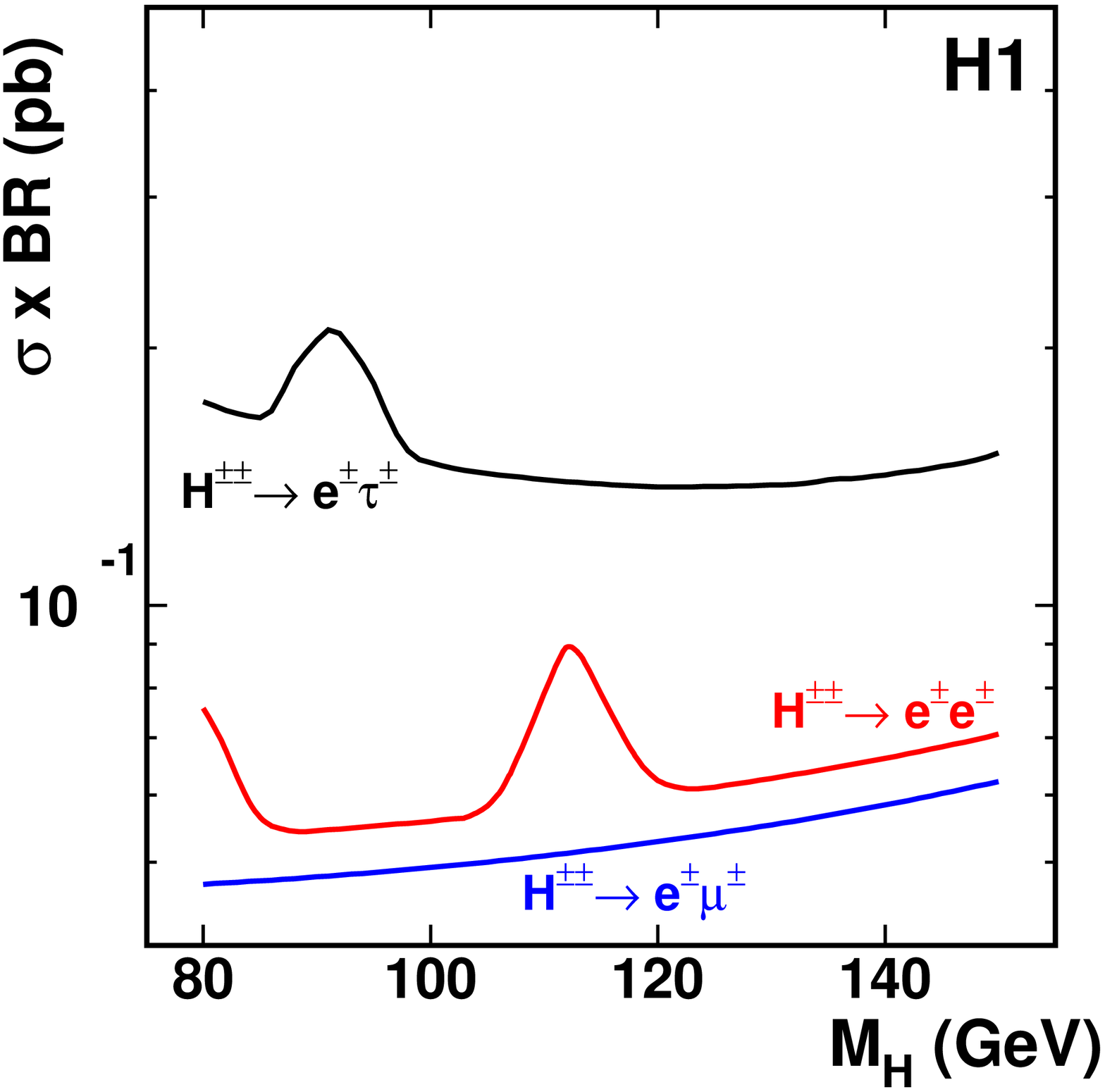,width=0.48\textwidth} 
\put(-65.,67.){{\large \bf (a)}} &
%
 \epsfig{file=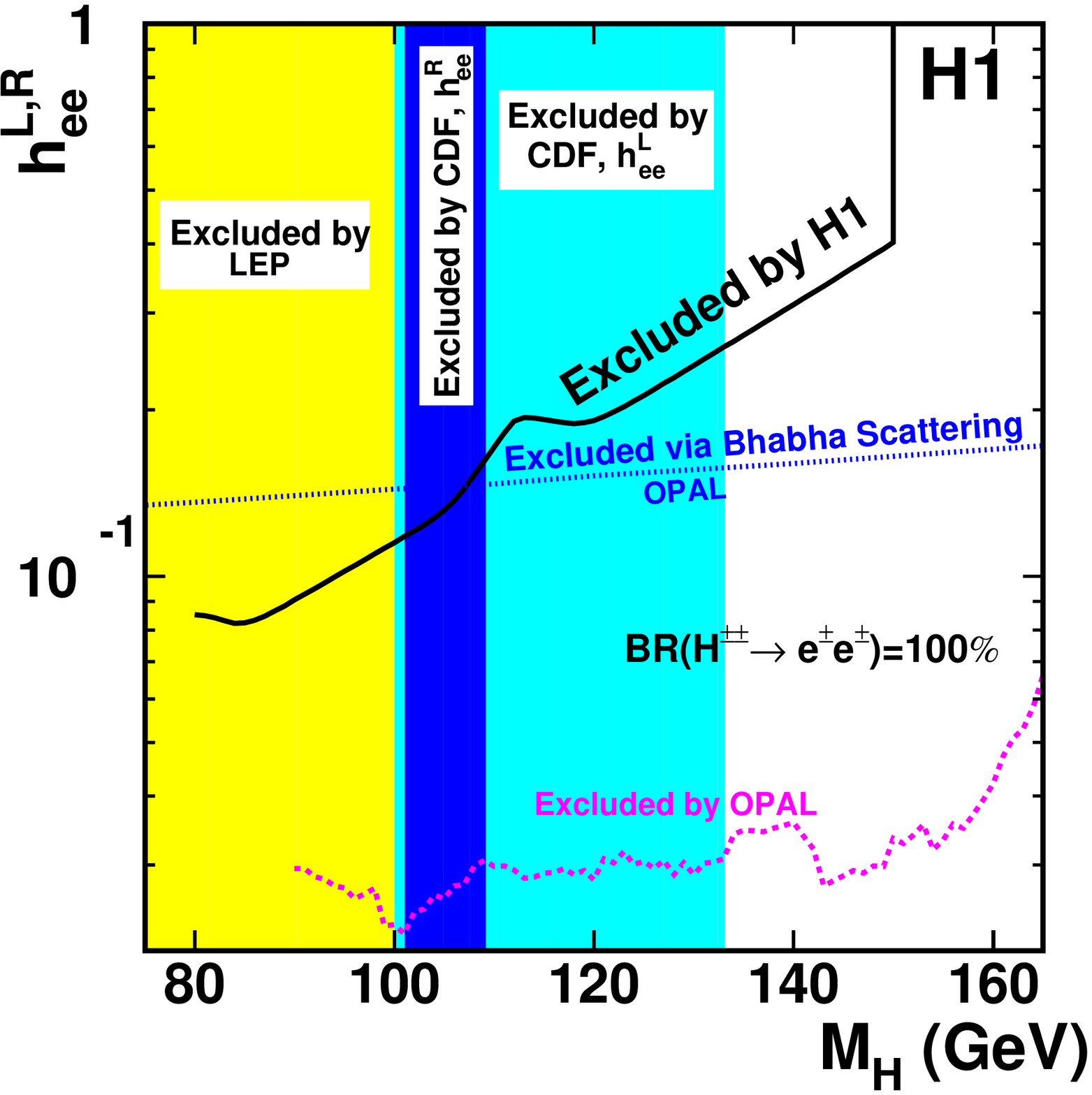,width=0.48\textwidth} \put(-65,67.){{\large \bf (b)}} \\
%
 \epsfig{file=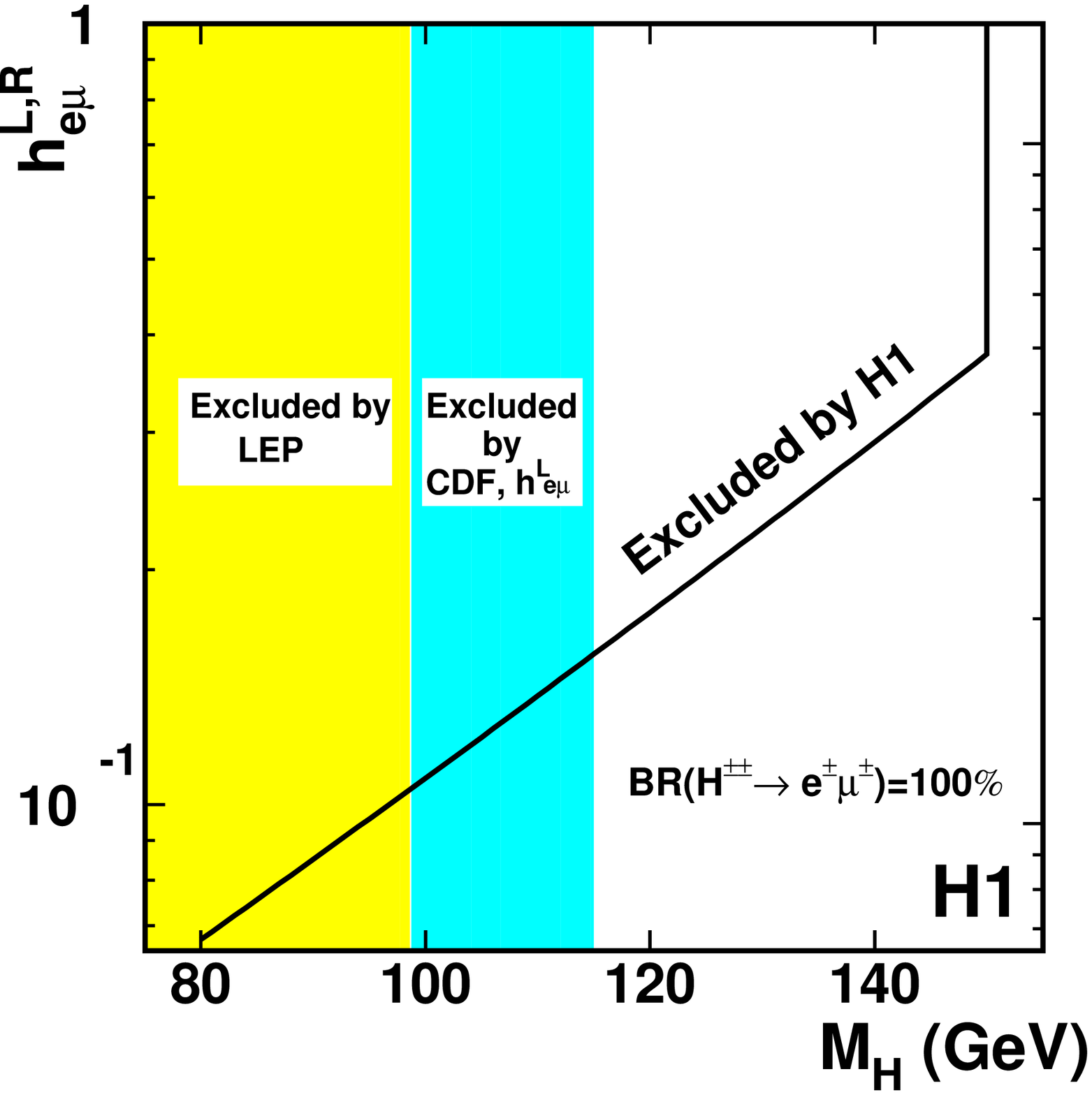,width=0.48\textwidth}\put(-65,67.){{\large \bf (c)}}  &
 \epsfig{file=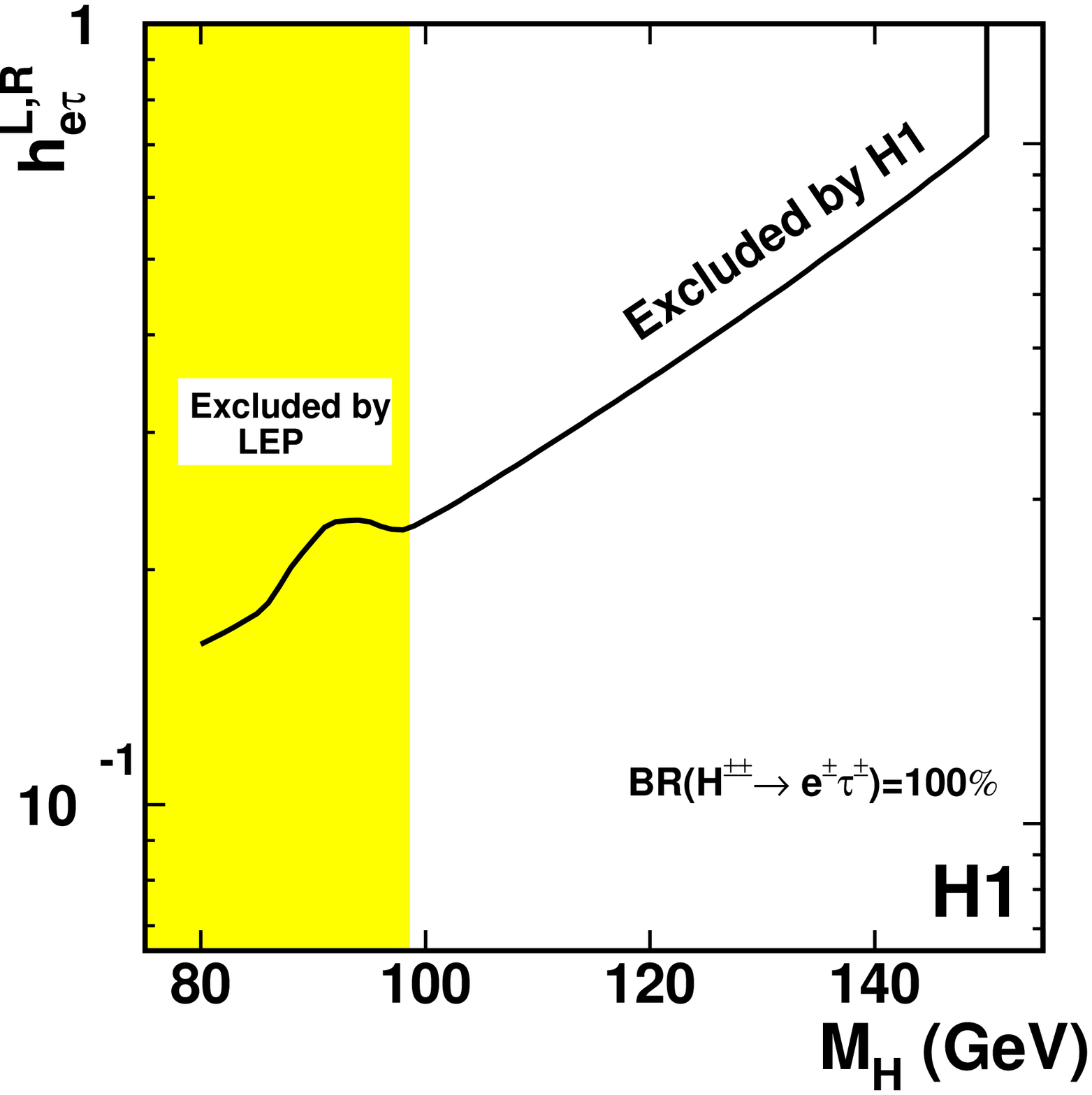,width=0.48\textwidth}\put(-65,67.){{\large \bf (d)}}
  \end{tabular}
  \end{center}
  \caption{
(a) Upper limits at the $95\%$ confidence level on the \hpp\ production
cross section times the branching ratio for the \hpp\ to decay into
$ee$, $e \mu$ or $e \tau$, as a function of the Higgs mass.
(b)-(d) Upper limits on the coupling $h_{el}$ assuming that the \hpp\
couples only (b) to $ee$, (c) to $e \mu$ or (d) to $e \tau$. 
Regions above the curves are excluded. The constraints obtained from
pair production at LEP and at CDF and from single production at OPAL
are also shown.
}
  \label{fig:limites}
\end{figure}
%

With the final Higgs selection no significant excess over the SM expectation
is observed.
Upper limits on the \hpp\ production cross section times the branching ratio for the 
\hpp\ to decay into one of the analysed final states are derived as a function of the
\hpp\ mass and are shown in Fig.~\ref{fig:limites}a.
The limits are presented at the $95\%$ confidence level and are obtained using a modified
frequentist approach~\cite{cl}.
Statistical uncertainties, as well as
the influence of the various systematic uncertainties on both the shape and the
normalisation of the mass distributions for signal and background events, are taken
into account. 
The best sensitivity is obtained for a \hpp\ produced and decaying via $h_{e \mu}$,
with upper limits around $0.05$~pb.

\noindent
Assuming that only one Yukawa coupling $h_{el}$ is non-vanishingly small,
these constraints are translated into mass dependent upper limits on the
coupling $h_{el}$, as shown in Fig.~\ref{fig:limites}b-d.

\noindent
If the doubly-charged Higgs boson couples only to an electron pair (Fig.~\ref{fig:limites}b)
the $ee$ analysis rules out \hpp\ masses below $138 \GeV$ for a coupling
$h_{ee}$ of the electromagnetic strength, $h_{ee} = 0.3$.
The result is compared to the bounds obtained from searches for \hpp\ 
pair production at LEP~\cite{leppairprod} and by the CDF experiment~\cite{cdf-pairprod},
and to both the indirect and direct limits obtained by the OPAL 
experiment~\cite{opal-singleprod}, the latter being the most stringent.
The OPAL experiment has also set similar stringent constraints on $h_{ee}$
independently of the Higgs decay mode. These constraints also exclude 
a sizeable \hpp\ production at HERA via $h_{ee}$ followed by the \hpp\ decay
via $h_{\mu\mu}$ or $h_{\tau\tau}$, which is consistent with the
non-observation of a resonance signal in the $\mu \mu$~\cite{mm} and
$\tau \tau$~\cite{ref:simon} final states in the present H1 data.

\noindent
Assuming that the doubly-charged Higgs boson couples only to an electron-muon
(electron-tau) pair, the $e \mu$ ($e \tau$) analysis allows masses below
$141 \GeV$ ($112 \GeV$) to be ruled out for
$h_{e\mu} = 0.3$ ($h_{e \tau} = 0.3$), as shown in Fig.~\ref{fig:limites}c 
(Fig.~\ref{fig:limites}d). 
The H1 limits extend the excluded region in the electron-muon and electron-tau channels
to masses that are beyond those reached in previous searches
for pair production at LEP~\cite{leppairprod} and the Tevatron~\cite{cdf-pairprod}.

\section{Conclusion}

A search for the single production
of doubly-charged Higgs bosons coupling to $ee$, $e\mu$ or $e\tau$ is presented.
In a previous model independent multi-electron analysis, H1 observed
six events with a di-electron mass above $100$~GeV,
a region where the Standard Model expectation is small.
Out of the six events, only one is compatible with the signature
of a doubly-charged Higgs boson.
No electron-muon or electron-tau event is found in this mass domain.

\noindent
This analysis places new limits on the \hpp\ mass
and its Yukawa couplings $h_{el}$ to an electron-lepton pair. 
Assuming that the doubly-charged Higgs boson only couples to electron-muon (electron-tau)
pairs, a limit of $141$~GeV ($112$~GeV) is obtained on the Higgs mass, for a coupling
$h_{e\mu} = 0.3$ ($h_{e \tau} = 0.3$) corresponding to an interaction of
electromagnetic strength.

\section*{Acknowledgements}

We are grateful to the HERA machine group whose outstanding
efforts have made this experiment possible. 
We thank
the engineers and technicians for their work in constructing and
maintaining the H1 detector, our funding agencies for 
financial support, the
DESY technical staff for continual assistance
and the DESY directorate for support and for the
hospitality which they extend to the non-DESY 
members of the collaboration.
We are especially grateful to J.~Maalampi and  N.~Romanenko for providing 
the doubly-charged Higgs Lagrangian implementation in CompHEP
which was 
used in this analysis.
We would also like to thank K.~Huitu and E.~Boos
for their help and valuable discussions.


\vspace{3cm}



\newpage

\begin{thebibliography}{99}
\bibitem{HTM} 
G.~B.~Gelmini and M.~Roncadelli,
Phys.\ Lett.\ B {\bf 99} (1981) 411.
%
\bibitem{pati}
J.~C.~Pati and A.~Salam,
Phys.\ Rev.\ D {\bf 10} (1974) 275;
R.~E.~Marshak and R.~N.~Mohapatra,
Phys.\ Lett.\ B {\bf 91} (1980) 222.
%
\bibitem{moha1} 
R.~N.~Mohapatra and G.~Senjanovic,
Phys.\ Rev.\ Lett.\  {\bf 44} (1980) 912.
%
\bibitem{moha3} 
G.~Senjanovic and R.~N.~Mohapatra,
Phys.\ Rev.\ D {\bf 12} (1975) 1502; 
%
R.~N.~Mohapatra and R.~E.~Marshak,
Phys.\ Rev.\ Lett.\  {\bf 44} (1980) 1316
[Erratum-ibid.\  {\bf 44} (1980) 1643].
%
%
\bibitem{LIGHT}
C.~S.~Aulakh, A.~Melfo and G.~Senjanovic,
Phys.\ Rev.\ D {\bf 57} (1998) 4174
[hep-ph/9707256];
%
Z.~Chacko and R.~N.~Mohapatra,
Phys.\ Rev.\ D {\bf 58} (1998) 015003
[hep-ph/9712359];
%
B.~Dutta and R.~N.~Mohapatra,
Phys.\ Rev.\ D {\bf 59} (1999) 015018
[hep-ph/9804277].
%
%
\bibitem{me} 
A.~Aktas {\it et al.}  [H1 Collaboration],
Eur.\ Phys.\ J.\ C {\bf 31} (2003) 17
[hep-ex/0307015].
%
\bibitem{zeus}
E.~Accomando and S.~Petrarca,
Phys.\ Lett.\ B {\bf 323} (1994) 212
[hep-ph/9401242].
%
%
%
\bibitem{indlim1} 
M.~L.~Swartz,
Phys.\ Rev.\ D {\bf 40} (1989) 1521;
%
J.~F.~Gunion {\it{ et al.}},
Phys.\ Rev.\ D {\bf 40} (1989) 1546;
%
M.~Lusignoli and S.~Petrarca,
Phys.\ Lett.\ B {\bf 226} (1989) 397;
%
G.~Barenboim, K.~Huitu, J.~Maalampi and M.~Raidal,
Phys.\ Lett.\ B {\bf 394} (1997) 132
[hep-ph/9611362];
%
S.~Godfrey, P.~Kalyniak and N.~Romanenko, Phys.\ Rev.\ D {\bf 65} (2002) 033009 [hep-ph/0108258].
%
\bibitem{opal-singleprod} 
G.~Abbiendi {\it et al.}  [OPAL Collaboration],
Phys.\ Lett.\ B {\bf 577} (2003) 93
[hep-ex/0308052].
%
\bibitem{DAVIDSON}
F.~Cuypers and S.~Davidson, Eur.\ Phys.\ J. C {\bf 2} (1998) 503.
%
\bibitem{VEGAS} G.P.~Lepage, CLNS-80/447 (1980).
%
\bibitem{comphep}
E.~Boos {\it{ et al.}} [CompHEP Collaboration], Nucl. Instrum. Meth. A {\bf 534} (2004) 250 [hep-ph/0403113]; A. Pukhov {\it{ et al.}}, ``CompHEP - a package for evaluation of Feynman diagrams and integration over multi-particle phase space", hep-ph/9908288; available at http://theory.sinp.msu.ru/comphep. 
%
%
\bibitem{ROMANENKO}
S.~Godfrey, P.~Kalyniak and N.~Romanenko,
Phys.\ Rev.\ D {\bf 65} (2002) 033009
[hep-ph/0108258].
%
\bibitem{cteq}
H.~L.~Lai {\it et al.},
Phys.\ Rev.\ D {\bf 55} (1997) 1280
[hep-ph/9606399].
%
\bibitem{pythia}
T.~Sj\"ostrand {\it{et al.}},
Comput.\ Phys.\ Commun.\  {\bf 135} (2001) 238
[hep-ph/0010017].
%
%
\bibitem{DGLAP}
V.~N.~Gribov and L.~N.~Lipatov,
Yad.\ Fiz.\  {\bf 15} (1972) 781
[Sov.\ J.\ Nucl.\ Phys.\  {\bf 15} (1972) 438];
G.~Altarelli and G.~Parisi,
Nucl.\ Phys.\ B {\bf 126} (1977) 298;
Y.~L.~Dokshitzer,
Sov.\ Phys.\ JETP {\bf 46} (1977) 641
[Zh.\ Eksp.\ Teor.\ Fiz.\  {\bf 73} (1977) 1216].
%
%
\bibitem{FORM}
 J.~A.~M~Vermaseren, ``New features of FORM", math-ph/0010025.
%
\bibitem{BRASSE}
 F.W.~Brasse {\it et al.}, Nucl.\ Phys.\  B {\bf 39} (1972) 421.
%
%
\bibitem{SOPHIA}
  A.~Mucke, R.~Engel, J.~P.~Rachen, R.~J.~Protheroe and T.~Stanev,
  Comput.\ Phys.\ Commun.\  {\bf 124} (2000) 290
  [astro-ph/9903478].
%
%
%
\bibitem{SLAC} 
R.~C.~Walker {\it et al.},
Phys.\ Rev.\ D {\bf 49} (1994) 5671.
%
\bibitem{dy} 
N.~Arteaga-Romero, C.~Carimalo and P.~Kessler,
Z.\ Phys.\ C {\bf 52} (1991) 289.
%
\bibitem{grape} 
T.~Abe,
Comput.\ Phys.\ Commun.\  {\bf 136} (2001) 126
[hep-ph/0012029].
%
%
%
%
%
\bibitem{django} 
 DJANGO~6.2;
 G.A.~Schuler and H.~Spiesberger,
 Proc. of the Workshop Physics at HERA,
 W.~Buchm\"uller and G.~Ingelman (Editors),
 Vol. 3 p. 1419 (October 1991).
%
\bibitem{wabgen} 
Ch.~Berger and P.~Kandel,
``A new Generator for Wide Angle Bremsstrahlung",
Proc. of the Monte Carlo Generators for HERA Physics Workshop, 
A.T.~Doyle, G.~Grindhammer, G.~Ingelman and H.~Jung (Editors),
DESY-PROC-1999-02, p. 596.
%
%
%
%
%
\bibitem{Abt:1996xv}
I.~Abt {\it et al.}  [H1 Collaboration],
Nucl.\ Instrum.\ Meth.\ A {\bf 386} (1997) 310 and 348.
%
\bibitem{Andrieu:1993kh}
B.~Andrieu {\it et al.}  [H1 Calorimeter Group],
Nucl.\ Instrum.\ Meth.\ A {\bf 336} (1993) 460.

\bibitem{h1calotestbeams}
B.~Andrieu {\it et al.}  [H1 Calorimeter Group],
Nucl.\ Instrum.\ Meth.\ A {\bf 344} (1994) 492;
{\it idem}, Nucl.\ Instrum.\ Meth.\ A {\bf 350} (1994) 57;
{\it idem}, Nucl.\ Instrum.\ Meth.\ A {\bf 336} (1993) 499.


\bibitem{spacal}
R.~D.~Appuhn {\it et al.}  [H1 SPACAL Group],
Nucl.\ Instrum.\ Meth.\ A {\bf 386} (1997) 397.



%

\bibitem{ref:simon} S.~Baumgartner, ``Search for doubly-charged Higgs decaying into $\tau$ leptons
at HERA'', PhD thesis, ETH Z\"urich, ETH-16125 (2005), available at
http://www-h1.desy.de/publications/theses\_list.html.

\bibitem{Adloff:2003uh}
C.~Adloff {\it et al.}  [H1 Collaboration],
Eur.\ Phys.\ J.\ C {\bf 30} (2003) 1 
[hep-ex/0304003].

\bibitem{mm}
A.~Aktas {\it et al.}  [H1 Collaboration],
Phys.\ Lett.\ B {\bf 583} (2004) 28
[hep-ex/0311015].
%


\bibitem{cl}  
T.~Junk,
Nucl.\ Instr.\ Meth.\ A {\bf 434} (1999) 435 [hep-ex/9902006].

%
\bibitem{leppairprod} 
J.~Abdallah {\it et al.}  [DELPHI Collaboration],
Phys.\ Lett.\ B {\bf 552} (2003) 127
[hep-ex/0303026];
P.~Achard {\it et al.}  [L3 Collaboration],
Phys.\ Lett.\ B {\bf 576} (2003) 18
[hep-ex/0309076];
G.~Abbiendi {\it et al.}  [OPAL Collaboration],
Phys.\ Lett.\ B {\bf 526} (2002) 221
[hep-ex/0111059].
%

\bibitem{cdf-pairprod}
D.~Acosta {\it et al.}  [CDF Collaboration],
Phys.\ Rev.\ Lett.\ {\bf 93} (2004) 221802
[hep-ex/0406073].



\end{thebibliography}
\end{document}